# em—ergence of the em—dash: a population-level rise in em-dash frequency in medRxiv preprints at the dawn of the large-language-model era

**Przemysław Czuma**

*Polish Association for Artificial Intelligence in Medicine*

## Abstract

**Background.** Large language models (LLMs) can leave subtle stylistic traces in text written with their assistance. One of the most frequently cited is the em-dash, the Unicode character U+2014. The common belief is that LLMs use this mark liberally, whereas human authors reach for it only sparingly. Human restraint may reflect typographic convention and habit, but also the simple difficulty of typing the character on a keyboard. Yet for all the anecdotes, no one has rigorously measured whether em-dash use has actually changed in the scientific literature in recent years. This is a period in which the rise of large language models may turn out to be one of the most visible marks left on written text.

**Methods.** The study, pre-registered in the Open Science Framework under the identifier HFT8C, was conducted on the full set of medRxiv XML files containing the full text of preprints from the official Text-and-Data-Mining resource. Initially, this comprised 82,067 first-version preprints. The primary cohort consisted of first, original versions of preprints deposited on medRxiv between 2020 and 2025 that had an extractable Discussion section of at least 500 characters (N = 69,632). The primary binary endpoint was the presence of at least one em-dash in the Discussion section. The principal measure was the absolute change in the prevalence of this character between the “pre-ChatGPT” era (before 30 November 2022) and the “post-ChatGPT” era. Estimation used a logistic model with standard errors robust to clustering by first author. The analysis plan, which also included six supporting analyses, six sensitivity analyses, and two falsification tests, was frozen before any confirmatory result was computed.

**Results.** The prevalence of the em-dash in Discussion sections rose from 4.23% before the public release of ChatGPT to 11.58% afterward. This corresponds to an absolute increase of 7.35 percentage points (95% CI 6.94–7.77; odds ratio 2.96, 95% CI 2.77–3.17). The increase was not a sharp jump at the moment of ChatGPT’s release, but rather a gradual, delayed acceleration: prevalence remained near 4% through 2023, reached 8.0% in 2024, and 20.3% in 2025. The effect survived every feasible sensitivity analysis (7.35–7.60 pp) and both falsification tests. A placebo temporal split within the pre-LLM era showed no meaningful change (+0.13 pp, 95% CI −0.33 to +0.58). In utility and boilerplate sections such as Acknowledgments and Data Availability, the effect was essentially absent. Independent LLM-associated lexical markers and within-paper comparisons of sections pointed in the same direction.

**Conclusions.** Across the full set of medRxiv preprints, em-dash use in the most narrative section of clinical work underwent a large increase with a distinctive timing, consistent with the growing spread of generative-AI writing tools. The marker is a population-level indicator, not a per-paper detector of LLM use: it cannot establish whether a given manuscript was produced with artificial intelligence, and the study design cannot establish causality. What it does show is that something in the way scientific literature is written changed markedly in the first half of the 2020s, and roughly when that change occurred.

---

# 1. Introduction

## 1.1 On the em-dash

Punctuation marks rarely draw attention. A comma, a full stop, an exclamation mark, or a question mark carries no specific content of its own. Its job is to organize text and make the meaning of written words clear.

The em-dash, here understood as the Unicode character U+2014, is, in this respect, an exception. It sets off a phrase with more drama than a comma or a hyphen, yet more subtly than parentheses. It also has no key of its own. On most keyboards, it has to be produced through a shortcut, an autocorrect rule, or a menu that many users never open.

For a human writer, using the mark is therefore often a deliberate act. For a large language model, by contrast, it is simply another token. A model needs no keyboard to produce an elegant flourish. It outputs the em-dash as readily as any other punctuation mark, and, as many users have noticed, it does so freely. This may follow from training on heavily edited professional prose, from mechanisms that reward the clear setting-off of ideas, or from both at once.

Over the past few years, the em-dash has travelled from a rather niche typographic ornament to an informal hallmark of text associated with LLMs: a mark that increasingly prompts readers to ask whether a text was written by a human alone and, if not, to what degree.

This remains anecdote. The present work sets out to test whether the data support it.

## 1.2 From story to measurement

The claim that “LLMs love em-dashes” is widespread in common opinion. Yet it has rarely been tested at scale in the actual scientific literature, and, to the author’s knowledge, essentially never with the rigour of a pre-registered confirmatory study.

One informal analysis had already pointed this way: Keck (2025) reported that the em-dash roughly doubled in ecology abstracts retrieved from OpenAlex between 2021 and 2025 [1]. Machine-modified text has been found elsewhere in the scholarly record as well; Liang et al. (2024), for example, documented LLM-modified content at scale in conference peer reviews [2]. What was missing was a confirmatory test of the em-dash signal itself, fixed in advance and read from a source that preserves the author’s original typography.

The most obvious place to look might seem to be large, openly available databases of published scientific articles. The natural first choice was PubMed, the largest online collection of medical publications. Preliminary searches, however, did not yield the expected results. Literally: zero em-dashes. Further exploration (Czuma, 2026) revealed that texts published on PubMed undergo typographic normalisation; a pre-registered audit of four APIs confirmed this quantitatively, with PubMed preserving typographic punctuation in only about 0.6% of abstracts [3]. Asking PubMed about em-dashes is therefore like asking a translator to preserve the spelling mistakes of the original: one can ask, but the system is built to do something else.

A better source proved to be medRxiv, a preprint server in the health sciences. Full texts are available as JATS XML files through the official Text-and-Data-Mining channel. Crucially for the present purpose, the XML preserves the author’s original typography, including the em-dash as a faithful character, encoded as the entity `—`.

Of all the parts of a scientific paper written according to the IMRaD scheme, the Discussion is by design the most narrative. It is where authors interpret results, speculate, weigh limitations, and persuade, rather than merely report dry facts. One trait of early LLMs such as ChatGPT-3 was their ability to tell a story fluently, though not always one that stayed close to the facts. If LLM-assisted free composition were to leave stylistic traces anywhere, the Discussion section seems a particularly likely place to find them. An added advantage of the em-dash as such a trace is its expected low baseline in the pre-LLM era, which should make any change in frequency clearly detectable against the noise.

### 1.3 What this study is, and is not

This is a confirmatory study with a deliberately narrow question, about a single punctuation mark, and with equally deliberate restraint in what it claims.

Earlier exploratory work by the author suggested a rise in em-dash use in medRxiv preprints. But exploratory findings carry limited weight, and the history of science is full of patterns that vanished once someone looked again under rules fixed in advance.

For that reason, the entire analysis plan (cohort, endpoint, model, every supporting and sensitivity analysis, the falsification tests, and the thresholds for interpreting effect size) was frozen and deposited in the Open Science Framework.

It should be stated plainly that the aim of this work was not to build a new artificial-intelligence detector. A single paper full of em-dashes may be the work of a person who simply likes to use them. A paper without a single em-dash may have been written entirely by a machine instructed to avoid them, or may have had them stripped out after the text was generated.

The study cannot identify which particular manuscripts were produced with LLM involvement, nor can it prove that LLMs caused the observed change. It shows something narrower but still important: whether, and roughly when, em-dash use changed across tens of thousands of preprints, and whether that change is temporally coincident with the growing popularity of LLMs as a working tool.

---

## 2. Methods

The full analysis plan was registered in the Open Science Framework (doi:10.17605/OSF.IO/HFT8C). The analysis code was frozen in a GitHub repository before the confirmatory analysis was run, at commit 21b65fd; the handling of publication dates later required corrections, described in §4.6.

What follows is a faithful summary of that plan. Deviations from it (two, both corrected before publication) are documented in §4.6.

## 2.1 Data source: why medRxiv and why XML

The complete full-text medRxiv corpus was retrieved on 26 May 2026 from the server's official Amazon S3 Text-and-Data-Mining resource, in JATS XML format [4]. It comprised 82,192 records: 82,067 first-version preprints and 125 later-version records.

XML was used rather than public HTML or any derived index for one essential reason: the XML preserves the em-dash as an actual character. Pilot exploration confirmed that the pipeline counts only the literal codepoint U+2014 and does not confuse it with similar marks that abound in scientific text: the ASCII hyphen U+002D, the Unicode hyphen U+2010, the en-dash U+2013 (used among other things for ranges), or the minus sign U+2212.

In our data, "—" means an em-dash and nothing else.

This also separates the present work from earlier em-dash observations drawn from metadata indexes such as OpenAlex, where the character does not always survive (the separate report mentioned above); here the mark that is measured is the mark the author entered [1].

It should be explained that on medRxiv the first published object is a PDF generated from the author's original manuscript, and that the full-text HTML and XML are produced from it by automated conversion, usually within 1–2 days, and within 48 hours according to the medRxiv Frequently Asked Questions [5]. The analysed XML is therefore a derived conversion of the submission, not the author's raw source file. What matters, however, is that, unlike the typographically normalised medRxiv HTML and PubMed, it preserves the em-dash as the literal entity U+2014, which was verified. Any influence of the conversion pipeline itself on typography was excluded by the utility-section placebo test (§3.4): had the conversion systematically inserted or transformed em-dashes, their frequency would also have risen in Acknowledgments and Data Availability, which it did not.

## 2.2 Cohort

The primary cohort consisted of first-version preprints, with deposit dates between 1 January 2020 and 31 December 2025, containing an extractable Discussion section at least 500 characters long, excluding tables and other non-prose constructs. The 500-character threshold corresponds to roughly one or two paragraphs of prose: enough to speak of a real Discussion rather than merely its beginning. The deposit date was derived from the medRxiv DOI in the format YYYY.MM.DD, not from the default date field in the raw dataset; details and rationale are given in §4.6.

This selection yielded 69,632 preprints.

The first deposited version of each preprint was used, rather than the latest, so that the analysed text would correspond to the authors' original submission, before any later revision could distort the dating.

The 500-character threshold excluded 79 preprints whose Discussion was shorter. A further 6,323 papers had no identifiable Discussion at all; almost all were short reports, letters, or non-standard formats in which a Discussion was genuinely absent. The exclusions were nearly symmetric across years.

The first seven months of medRxiv's existence, that is the year 2019, and the first four months of 2026, from January to April, fell outside the six full calendar years covered by the study and were set aside as supplementary cohorts.

## 2.3 Endpoint and exposure

The primary endpoint was binary: does the preprint's Discussion section contain at least one em-dash?

Presence or absence of the mark was chosen deliberately over the number of em-dashes in the section. It was judged to be the simplest and most robust version of the question: verifiable at a glance by the reader, and robust to a single paper that happens, for whatever reason, to contain dozens of em-dashes, for example in range descriptions, formulae, or unusually formatted passages. The number of em-dashes in the section was analysed separately as a supporting analysis.

The exposure was era. The dividing point was 30 November 2022, the day of ChatGPT's public release [6]: the first moment at which a capable generative model became freely available to essentially anyone and entered general awareness. A preprint deposited on or after that day was classified as "post"; an earlier one as "pre".

No wash-out window was applied. A preprint was counted by the date on which it was actually deposited.

## 2.4 Primary analysis and the agreed way of reading it

The primary test was a logistic regression of em-dash presence on era, with standard errors clustered by first author [7].

The clustering matters: prolific authors may have deposited many papers during this period, and their personal stylistic habits (a fondness for the em-dash or its consistent avoidance) should not be counted as many independent voices.

First authors were identified by ORCID where available, which covered 76.8% of the cohort. In the remaining cases, a conservative key combining surname, initial, institution, and country was used.

The reporting hierarchy was fixed in advance, so that no number could be chosen after the fact because it most flattered the expected result. The primary estimand was the absolute difference in em-dash prevalence between eras, expressed in percentage points. The prevalence ratio [8] and the odds ratio were reported alongside as relative measures, with the understanding that the odds ratio may overstate the intuitive sense of the effect at such a low baseline.

The interpretation of the result was also fixed in advance: an absolute increase of at least 5 percentage points was to be called "substantial", an increase of 2–5 points "moderate", and an increase below 2 points "trivial", regardless of statistical significance. With a corpus this large, almost any non-zero difference is statistically significant. Fixing the effect-size scale in advance prevents the word "significant" from impersonating the word "important".

For the record: the smallest difference the study could reliably detect was about 0.4–0.5 percentage points. The question was therefore never whether the effect could be seen, but whether it was large enough to matter.

## 2.5 Supporting, sensitivity, and falsification analyses

Six supporting analyses were pre-registered to view the same phenomenon from different angles:

- the number of em-dashes per Discussion, modelled as a length-adjusted rate [9];
- a within-text comparison of the Discussion with the Results;
- a check of eight strongly over-represented, LLM-associated lexical markers from Kobak et al. [10] in abstracts;
- prevalence year by year;
- an interrupted-time-series model testing whether the trajectory changed at the point of ChatGPT's release [11];
- subject-area composition across the two eras.

The marker list was corrected before publication, as described in §4.6.

Six sensitivity analyses repeated the primary test on modified cohorts: after excluding the small overlap with the author's earlier exploratory sample; after excluding COVID-19 papers; with stricter and looser length thresholds; after excluding papers whose Discussion was fused with the Results or Conclusion section; and after reclassifying a small group of papers in which the only em-dash occurred within a numeric range, for example "1934—1971".

A seventh planned check, repeating the analysis on the latest versions rather than the first, could not be carried out because the frozen extraction, by design, covered only first versions. This is reported as a limitation.

The two falsification tests were an important and deliberately risky part of the plan: analyses capable of overturning the main interpretation of the result.

The first was a placebo-breakpoint test. Only preprints from the pre-LLM era were analysed, from 2020 to November 2022, that is, from a period in which ChatGPT could not plausibly explain a change. These were split at an arbitrary midpoint, 30 June 2021. If em-dashes were simply drifting upward for reasons of fashion or editorial style, this false breakpoint should also show an increase.

The second test was a utility-section placebo: counting em-dashes in parts of the paper that authors are unlikely to polish with an LLM, namely the Acknowledgments and the Data Availability statement. If the increase were a platform-wide composition artifact (for example, the result of applying modern editing tools to the whole text rather than a change in writing at the narrative level), it should appear there too.

## 2.6 Measurement validation

Before the analyses were run, the pipeline was checked manually against the source XML on two pre-registered samples of 200 preprints each. The first served to confirm that the automatically identified Discussion section was the correct section and had the correct boundaries. The second, enriched with rare positive cases, served to confirm the correctness of em-dash counting.

Because the object being checked is unambiguous (a specific character is either present or absent), agreement was measured against a human-verified ground truth.

The result was perfect: section identification agreed in 200 of 200 cases, and em-dash-presence classification likewise in 200 of 200 cases. Sensitivity and specificity were 100%, and Cohen’s kappa was 1.0.

## 2.7 Software and reproducibility

The analyses were performed in Python. The section-comparison model used generalised estimating equations (GEE) as a fallback to the pre-registered mixed model, which was retained as the primary specification.

All code is public and frozen at the commit indicated above. Processed per-paper measurements are archived on Zenodo (doi:10.5281/zenodo.20557419), released at publication.

As a reproducibility check, all confirmatory, supporting, and sensitivity analyses were independently re-implemented from scratch, from the pre-registered specification and the frozen dataset, blind to the reported estimates; they reproduced the stated values. In addition, a random sample of papers was compared against the raw JATS XML files, confirming faithful extraction of the em-dash (U+2014), the DOI-derived deposit date, and the version number.

The medRxiv corpus itself is public external data and is not redistributed here.

# 3. Results

## 3.1 The main result

Across the 69,632 preprints of the primary cohort, the proportion of Discussion sections containing at least one em-dash rose from 4.23% before the public release of ChatGPT to 11.58% afterward. This corresponds to a rise from 1,368 of 32,324 preprints to 4,322 of 37,308.

This was an absolute increase of 7.35 percentage points, with a 95% confidence interval from 6.94 to 7.77 points. The result was clearly above the pre-set “substantial” threshold of 5 points and roughly sixteen times the smallest effect the study could detect.

The odds ratio was 2.96, with a 95% confidence interval from 2.77 to 3.17; the prevalence ratio was 2.74.

The p-value, equal to $2 \times 10^{-222}$, is given for completeness. At this sample size, it is the least interesting number in this paragraph.

The first-author clustering that produced these intervals covered 53,430 distinct authors. The median was one preprint per author, and the maximum was 47. This means that the honest, dependence-adjusted sample was large, though not as large as the raw paper count might suggest. The confidence intervals already reflect this.

## 3.2 Not a step, but a take-off

The more interesting story lies in the dating.

If ChatGPT's release had acted like a switch, prevalence should have jumped sharply at the start of 2023. It did not. For three quarters, the baseline held quietly around 4%. It then doubled in 2024, reaching, in the third quarter of 2025, almost a quarter of preprints (23.5%).

The pre-registered quarterly interrupted-time-series model, covering 24 quarters with a breakpoint at the date of ChatGPT's release, tells the same story in its coefficients: essentially no prior trend, with a pre-breakpoint slope $\beta_1 = 0.010$ per quarter, $p = 0.27$, and the effect concentrated in a sharp positive change of slope after the breakpoint, $\beta_3 = +0.199$, 95% CI 0.178–0.220, rather than in an immediate level jump, for which $\beta_2 = -0.80$.

The trajectory is shown in Figure 3.

Whatever happened, it happened gradually, and then quickly. The shape of the curve is more consistent with general societal adoption of a new tool and the maturing of the technology in 2024–2025 than with overnight enthusiasm following the launch of a breakthrough product.

The two supplementary cohorts, outside the registered 2020–2025 window, bracket this trajectory. In 2019, medRxiv's first and partial year, prevalence was 5.98% (n = 869), close to the pre-LLM baseline; in the first four months of 2026 it was 20.89% (n = 4,845), in line with the high 2025 level.

## 3.3 The same signal from six sides

Every supporting analysis pointed in the same direction.

The number of em-dashes per Discussion, adjusted for text length, rose 3.3-fold. The incidence rate ratio was 3.33, with a 95% confidence interval from 3.05 to 3.63.

Eight LLM-associated style words from Kobak et al. [10], vocabulary unrelated to punctuation, rose in parallel in the abstracts. The composite odds ratio was 4.05. This was an independent fingerprint, yet it moved in the same direction and at the same time.

The em-dash itself, measured in the same way at the abstract level, just as the words were measured in Kobak, had an odds ratio of 2.92, with a 95% confidence interval from 2.60 to 3.27. This was nearly identical to the odds ratio for the em-dash in the Discussion, which was 2.96.

Within individual papers, the em-dash rose more in the Discussion than in the Results. The section–era interaction had $p = 2 \times 10^{-6}$. This is an asymmetry consistent with the premise that narrative prose is the most likely place for em-dashes generated or introduced with LLM involvement to concentrate. With a mechanical, whole-text cause such as a word processor, the distribution across sections should be more even.

## 3.4 A failed attempt to overturn the result

An effect with an elegant story should always invite suspicion. That is why the falsification tests counted almost as much as the result itself. Both came out clean.

The placebo breakpoint, placed within the pre-LLM era where no real ChatGPT-related effect should exist, gave an absolute change of +0.13 percentage points, with a 95% confidence interval from −0.33 to +0.58. It was indistinguishable from zero.

The pre-LLM scientific literature was therefore not quietly drifting upward in em-dash use. The increase was specific to the period after generative models became available, not to a fashion that merely happened to coincide with the date of their release.

The utility-section placebo was likewise reassuring. In Acknowledgments, the em-dash frequency moved from 0.22% to 0.46%, an increase of +0.24 percentage points. In Data Availability statements it rose from 0.000% to 0.003%.

It is hard to assume that authors widely used LLMs to smooth standard Data Availability statements, and these data tend to agree. A platform-wide composition shift should have increased em-dashes in these boilerplate sections too. Yet over the study period, essentially no such increase occurred.

## 3.5 Robustness

The primary effect was robust to how the cohort was defined.

Across the five feasible sensitivity analyses, the absolute increase stayed within a narrow range from 7.35 to 7.60 percentage points, against a primary value of 7.35: after excluding the exploratory-sample overlap it was 7.37; after excluding COVID-19 papers, 7.60; with the length threshold tightened or loosened, 7.35 and 7.36; after excluding Discussions fused with other sections, 7.43; and after reclassifying 36 papers in which the only em-dash occurred within a numeric range, 7.37.

The sixth planned check, the latest-version analysis, was infeasible on a frozen dataset containing only first versions and was recorded as a limitation, not a result.

## 3.6 Exploratory: self-declared AI use and the matter of the em-dash

The following analysis was not part of the frozen plan and is reported as exploratory, added after registration. It asks a more direct question than the rest of the work permits: whether preprints that themselves declare the use of a generative model in preparing the manuscript contain an em-dash in the Discussion more often than those that make no such declaration. To hold time constant, comparisons were made within individual years of the post-ChatGPT era; raw proportions with 95% Wilson confidence intervals [12] are given, with no comparative test.

Declarations of AI use were essentially absent before 2023, after which their share rose: 0.53% of 2023 preprints (95% CI 0.41–0.70), 1.22% of 2024 preprints (95% CI 1.03–1.43), and 3.61% of 2025 preprints (95% CI 3.32–3.93). Placed on a single time axis, the share of declarations and the frequency of the em-dash rise in parallel after ChatGPT's release (Figure 5): two independent signals of the same change, one self-reported and one typographic.

In each of these years, preprints with a declaration showed a higher frequency of em-dashes in the Discussion than the rest; for example, 32.1% (95% CI 28.2–36.3) versus 19.9% (95% CI 19.2–20.5) in 2025. Tellingly, the elevation was confined essentially to papers declaring AI use: the handful of papers that explicitly declared no such use ($n = 57$ in 2025; 24.6%, 95% CI 15.2–37.1) did not differ from the non-declaring majority (about 20%), instead of falling to a lower, “purely human” level. The pattern is therefore two-tiered (declared AI use clearly higher, everything else at a common baseline) rather than a graded gradient.

The caveats are the same as before and keep this analysis strictly exploratory. The declaration is voluntary and still rare, so the non-declaring group mixes genuine non-users with users who did not disclose.

AI declarations were detected by an automated detector whose accuracy was assessed by an independent language model acting as a judge (Claude Haiku, n = 2,162): precision 79.6%, recall 82.0%; the per-item judgments are made available in a separate file. The detector is therefore meant to distinguish a disclosure of writing assistance from papers whose subject is the language model itself, but it does so with moderate accuracy, which biases the observed contrast further toward the null, making it, if anything, an underestimate. This is neither a per-paper test nor proof of causation: most declaring papers still contain no em-dash, and some non-declaring papers do. It is one further independent, self-reported signal of LLM use set alongside the typographic marker, strengthening the plausibility of the interpretation without confirming it in any confirmatory sense.

---

# 4. Discussion

## 4.1 What the study showed

In a complete, population-level corpus of clinical preprints, em-dash use in Discussion sections roughly tripled after the public arrival of generative language models, reaching, in 2025, almost one in five preprints, and in the third quarter of that year almost one in four (23.5%).

The increase was “substantial” by the pre-set threshold, robust to every feasible sensitivity analysis, and, most importantly, survived analyses designed to explain it away as spurious. The placebo breakpoint within the pre-LLM era showed no effect. The boilerplate sections of the same papers barely moved. Two independent fingerprints, em-dash count and LLM-preferred vocabulary, rose in concert. The exploratory, post-hoc analysis of declared AI use as a writing aid lined up in the same direction.

When several independent cues lean together, the simplest reading is that they are responding to the same underlying change.

## 4.2 What it means, and the discipline of saying nothing more

It would be tempting to read the result “20% of 2025 preprints contain an em-dash” as “ChatGPT wrote 20% of preprints”. This temptation must be firmly resisted.

The em-dash is a population-level marker, not a per-paper detector. Some people love em-dashes. Some LLM users delete them. The mark decides nothing about any single manuscript, and no such attempt is made in this work. The frequency of a certain typographic habit rose across tens of thousands of documents, not the number of proven cases of machine authorship.

An observational design of this kind also cannot establish causation. What can be stated precisely is narrower and, in my view, still important: a large, abrupt change in the way clinical preprints are written appeared at almost exactly the same time as LLM-assisted writing became widely available, and is consistent with that explanation. The phrase “caused by” is deliberately not used here.

Moreover, LLM use does not reduce to a simple division between text written by a human and text written by a machine. In practice, it is a continuum. At one end is full text generation from a short prompt; further along are editing, shortening, smoothing style, translating, ordering the argument, or giving the text a more professional tone.

For this reason, the presence of an em-dash is not a machine author’s signature. It may be the trace of editorial help, translation, stylising, a human habit, or mere chance. That is exactly why this work is concerned not with the single manuscript but with the shift that becomes visible only at the level of a whole population of texts.

## 4.3 Why an acceleration, not a jump

The very shape of the curve is telling. If the sudden free access to ChatGPT at the end of 2022 explained the whole story, we would observe a sharp jump in 2023. Instead, 2023 was almost indistinguishable from the pre-LLM baseline, and the real movement came only in 2024 and 2025.

This delayed acceleration fits a familiar pattern: the slow diffusion of a new technology through a cautious community of professionals. Onto this is layered a second factor: the LLMs themselves improved, producing prose good enough that researchers began to entrust their own texts to them, for interpretation, smoothing, ordering, or translation.

Free availability was probably a necessary condition, but it does not explain the pace of the change. Accounting for the dynamics of technology adoption and the maturing of the tools does so much better.

The data, however, do not allow these factors to be separated. I therefore offer this interpretation as the reading most consistent with the trajectory, not as a demonstrated mechanism.

## 4.4 Why this matters beyond a typographic curiosity

When a person uses an em-dash instead of another mark, it is a small thing. When a similar trace appears en masse in scientific text during the rapid spread of large language models, the implications may be broader.

Preprints are prototypes of scientific papers, a large share of which will later reach the pages of journals. Moreover, this form of publication is gaining ever greater standing in its own right: it is increasingly read, cited, and incorporated into systematic reviews and guidelines long before formal peer review.

This does not mean that medRxiv is a simple model of the whole scientific literature. It does mean that it is a useful, well-dated, and technically accessible place to observe a change in writing style. If the way preprints come into being is changing under the influence of generative artificial-intelligence tools, that at least deserves to be noted.

How scientific information is produced may affect the trust placed in it. It therefore becomes a question of transparency, of disclosing when tools were involved, and of the norms we want to build around machine-assisted science. These are concerns common to all of science, not only to medicine, and certainly not only to medRxiv. medRxiv is simply the place where the typography survived intact, making the measurements presented here possible.

The clinical context undoubtedly raises the stakes, because this is literature that feeds real medical decisions. The phenomenon itself, however, does not belong to medicine alone. A parallel rise in em-dash use has been reported in ecology abstracts (Keck 2025 [1]), which suggests the shift is not specific to clinical writing.

## 4.5 Limitations

Several limitations deserve emphasis.

First, and most important, the em-dash is not a detector. The marker carries no meaning at the level of a single paper and does not justify any causal claim, as emphasised throughout.

Second, one pre-registered sensitivity analysis, concerning latest versions, could not be carried out because the frozen dataset contains only first versions. This is reported, not concealed.

Third, there is no clean typographic negative control, because in this corpus essentially all common punctuation drifts slightly upward over time. Specificity was therefore addressed differently: through the placebo breakpoint, the within-text section comparison, and the independent lexical markers. None of these, alone or together, proves that LLM use is the sole cause.

Fourth, clustering by first author does not capture every form of authorial dependence. Senior authors and laboratories may also shape the style of many papers.

Fifth, the corpus comes from a single preprint server in the health sciences. Whether the same curve appears in other areas of the literature remains an open and testable question, and may be regarded as the natural next step.

## 4.6 Deviations from the pre-registration

After the frozen plan was deposited, two deviations arose, and both were corrected before publication. Both are documented here; neither changes the substantive result.

The first concerned the handling of dates. After the code was frozen before analysis, at commit 21b65fd, work on the figures revealed two date-parsing problems. First, the deposit-date field initially defaulted the month and day to January, so that the registered daily cutoff, 30 November 2022, behaved in practice as a calendar-year cutoff. Second, medRxiv introduced a second DOI prefix, 10.64898, in December 2025, which the original parser did not recognise.

Both problems were resolved by deriving the deposit date from the medRxiv DOI for both prefixes. A parseable date was obtained for 99.1% of records; the rest concerned pre-format DOIs from 2019 and thus fell outside the cohort window. All date-dependent analyses were then recomputed: the primary contrast, the placebo breakpoint, the interrupted-time-series model, and all sensitivity analyses.

Relative to the frozen run, the primary estimate moved from +7.54 to +7.35 percentage points, and the cohort from 69,582 to 69,632 preprints. The substantive verdict, the placebo result, and all sensitivity and supporting analyses remained essentially unchanged.

No preprint in the primary cohort relies on a surrogate date. The pre-analysis extraction was preserved unchanged for audit, and the published dataset carries the corrected date column.

The second deviation concerned the lexical-marker list. This analysis tests markers of over-represented vocabulary from Kobak et al. [10] (2025).

The pre-registered list of eight words was compiled with LLM assistance. On verification against the published Kobak list, it turned out that two of them, *boast* and *underpinning*, did not appear on Kobak's list.

Keeping faith with the intent of the pre-registration, a set of markers was introduced that was derived by an objective rule taken wholly from Kobak: the eight style words with the highest over-representation in his published data were selected, after deduplication by word stem. The corrected set retains five of the original eight markers and replaces three.

Tellingly, when recomputed on the identical cohort, the composite odds ratio was essentially unchanged: 4.05 in abstracts and 4.08 in Discussions, against 4.38 and 4.24 respectively for the original list, with overlapping confidence intervals. The direction, significance, and convergence with the em-dash signal remained intact.

The original registration text was left unchanged, and the correction was recorded in a dated OSF addendum.

Fittingly for a paper about detecting machine-written text, the slip was a language-model error, and it was caught by a human author's return to the source.

## 4.7 Conclusions

The em-dash, for years a rather niche mark of typographic licence, became measurably more frequent in clinical preprints just as generative-AI tools came into common use.

Care was taken to show this carefully: within a frozen plan, with passing falsification tests and a pre-set interpretation of effect size.

A punctuation mark will not decide whether, or to what degree, a given paper was produced by a machine or a human. But it can say, across a whole corpus, that the change is real, large, and locatable in time.

And sometimes, in a record as vast and impersonal as the scientific literature, knowing that something shifted, and when, may be the first useful fact.

# Tables

## Table 1. Primary endpoint: em-dash presence in Discussion sections by era

Primary cohort: first-version medRxiv preprints deposited 2020–2025, with a Discussion ≥500 characters (N = 69,632). The era split was set at the public release of ChatGPT, 30 November 2022.

| Era | Preprints | With em-dash | Prevalence (95% CI) |
|---|---:|---:|---:|
| Pre-ChatGPT | 32,324 | 1,368 | 4.23% (4.02–4.46) |
| Post-ChatGPT | 37,308 | 4,322 | 11.58% (11.26–11.91) |
| Absolute difference | | | +7.35 pp (6.94–7.77) |

Prevalence ratio 2.74; odds ratio 2.96 (95% CI 2.77–3.17), with standard errors robust to clustering by first author; Wald $p = 2 \times 10^{-222}$. Pre-set effect-size scale: ≥5 pp as a "substantial" increase; the observed effect is about 16 times the minimum detectable difference (about 0.45 pp). Era was assigned by the DOI-derived medRxiv deposit date; see §4.6.

## Table 2. Em-dash frequency in Discussion sections by deposit year

Primary cohort.

| Year | Preprints | With em-dash | Prevalence (95% CI) |
|---|---:|---:|---:|
| 2020 | 12,061 | 482 | 4.00% (3.66–4.36) |
| 2021 | 11,367 | 524 | 4.61% (4.24–5.01) |
| 2022 | 9,681 | 387 | 4.00% (3.63–4.41) |
| 2023 | 10,281 | 454 | 4.42% (4.04–4.83) |
| 2024 | 12,097 | 971 | 8.03% (7.56–8.52) |
| 2025 | 14,145 | 2,872 | 20.30% (19.65–20.97) |

The baseline held near 4% through 2023, doubled in 2024, and reached one in five preprints in 2025. Year was assigned by the DOI-derived deposit date.

## Table 3. Sensitivity analyses

Primary estimand: absolute difference in percentage points. Concordance was defined in advance as the same direction of effect, a value within ±30% of the primary estimate, and overlapping 95% confidence intervals.

| Analysis | N | Difference (pp) | Concordant |
|---|---:|---:|---|
| Primary | 69,632 | 7.35 | reference |
| Exclude exploratory-sample overlap | 68,074 | 7.37 | yes |
| Exclude COVID-19 papers | 48,827 | 7.60 | yes |
| Discussion ≥250 characters | 69,702 | 7.35 | yes |
| Discussion ≥1,000 characters | 69,191 | 7.36 | yes |
| Exclude fused Discussion sections | 66,938 | 7.43 | yes |
| Reclassify range-only em-dashes | 69,632 | 7.37 | yes |
| Latest-version cohort | n/a | infeasible | n/a |

The latest-version analysis could not be performed, because the frozen extraction by design covered only first versions. This is reported as a limitation, not a result.

## Table 4. Convergent lexical markers in abstracts by era

The eight most over-represented style words from Kobak et al. (2025), pre- and post-ChatGPT; odds ratios from logistic models, as independent corroboration of the typographic signal. The marker list was corrected before publication; see §4.6. The parallel Discussion-section analysis is in the Supplement, Table S2.

| Word | Pre % | Post % | OR (95% CI) |
|---|---:|---:|---:|
| delve | 0.01 | 0.13 | 14.17 (4.42–45.46) |
| groundbreaking | 0.00 | 0.03 | 11.27 (1.47–86.14) |
| underscore | 0.73 | 4.60 | 6.55 (5.68–7.55) |
| intricate | 0.07 | 0.42 | 6.13 (3.85–9.75) |
| meticulously | 0.02 | 0.06 | 3.81 (1.44–10.07) |
| garner | 0.03 | 0.10 | 3.47 (1.67–7.20) |
| showcase | 0.11 | 0.25 | 2.33 (1.58–3.44) |
| leverage | 1.05 | 2.30 | 2.23 (1.96–2.54) |
| Any of the eight | 1.97 | 7.52 | 4.05 (3.70–4.44) |

# Figures

Final figure files, in PDF and PNG at 300 DPI (v3, with the dual-prefix DOI-corrected data), are in the directory results/figures_final/: fig1_per_year_prevalence, fig2_segmented_its_quarterly, fig3_kobak_abstracts_forest, fig4_declaration_vs_emdash, fig_study_flow.

## Figure 1

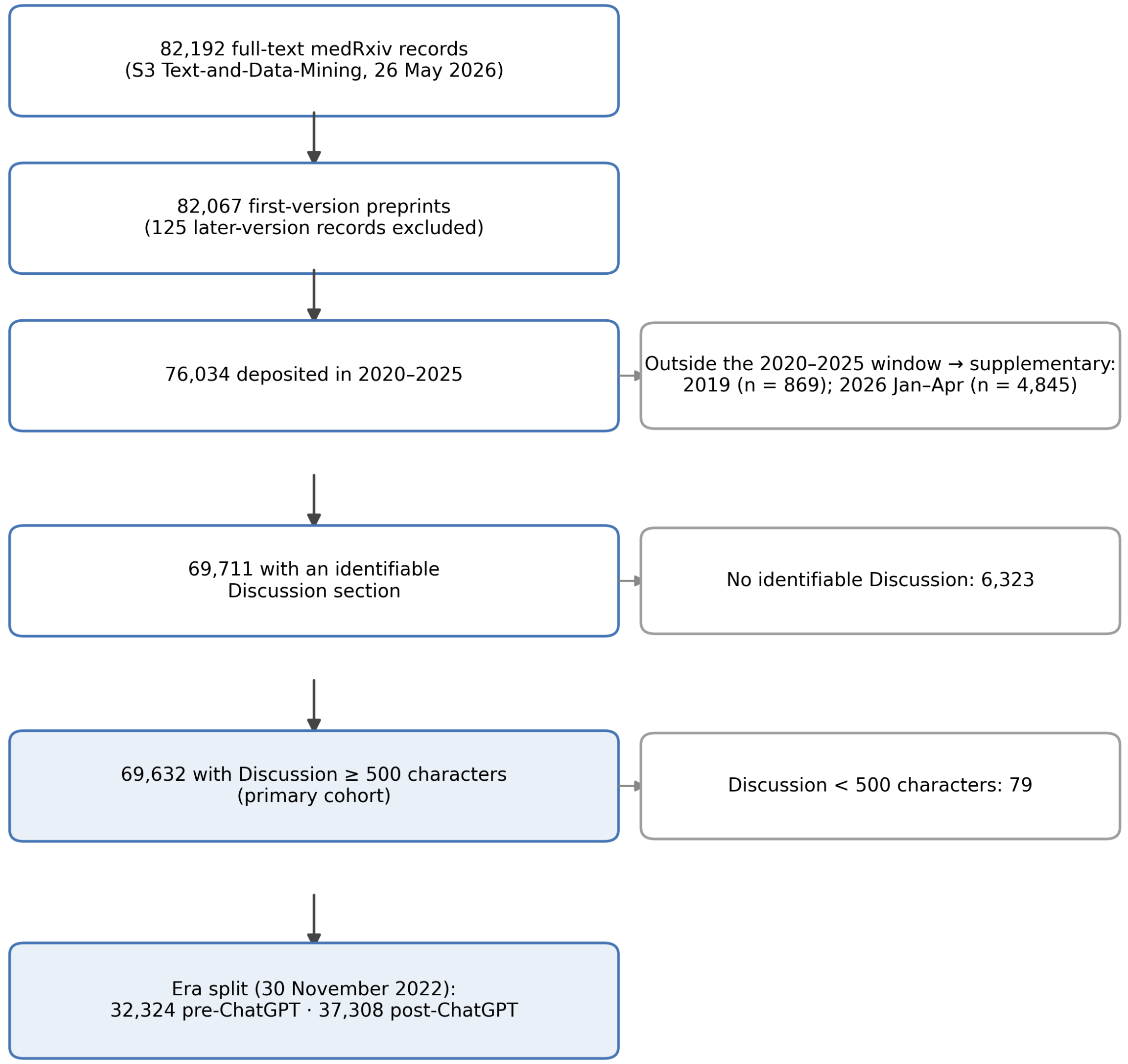

**Figure 1. Study flow: cohort selection.** From 82,192 full-text medRxiv records retrieved from the S3 Text-and-Data-Mining resource on 26 May 2026 to the primary cohort of 69,632 first-version preprints with a Discussion section of at least 500 characters: 32,324 pre-ChatGPT and 37,308 post-ChatGPT. Two periods outside the 2020–2025 window (the year 2019, n = 869, and January–April 2026, n = 4,845) were set aside as supplementary cohorts. File: fig_study_flow.pdf.

**Figure 2**

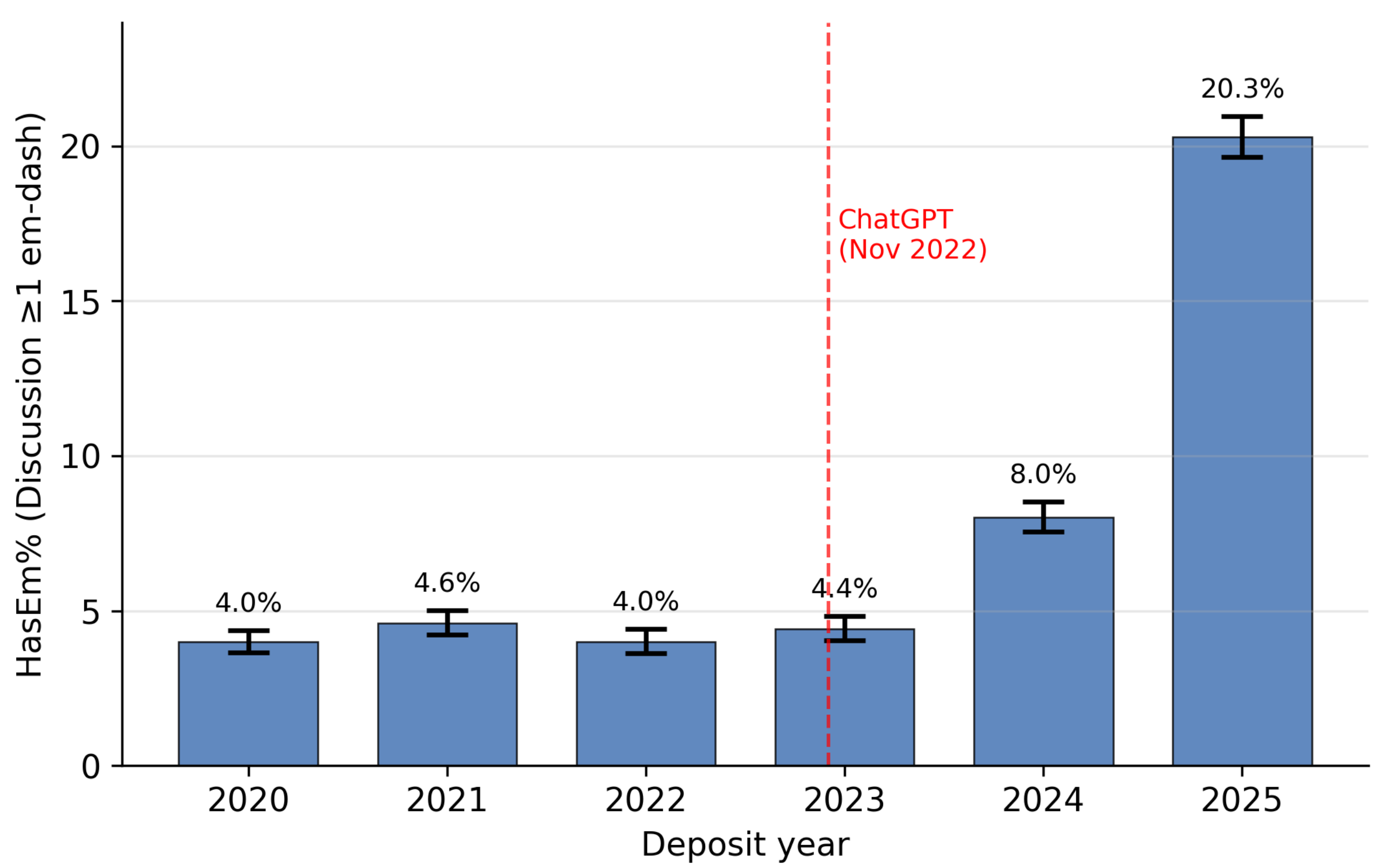

**Figure 2. Em-dash frequency in Discussion sections by deposit year, 2020–2025.** Bars show the percentage of first-version medRxiv preprints whose Discussion section contains at least one em-dash, with 95% Wilson confidence intervals. The dashed line marks the public release of ChatGPT in November 2022. Prevalence holds flat near 4% through 2023, doubles in 2024, and reaches 20.3% in 2025. File: fig1_per_year_prevalence.pdf.

## Figure 3

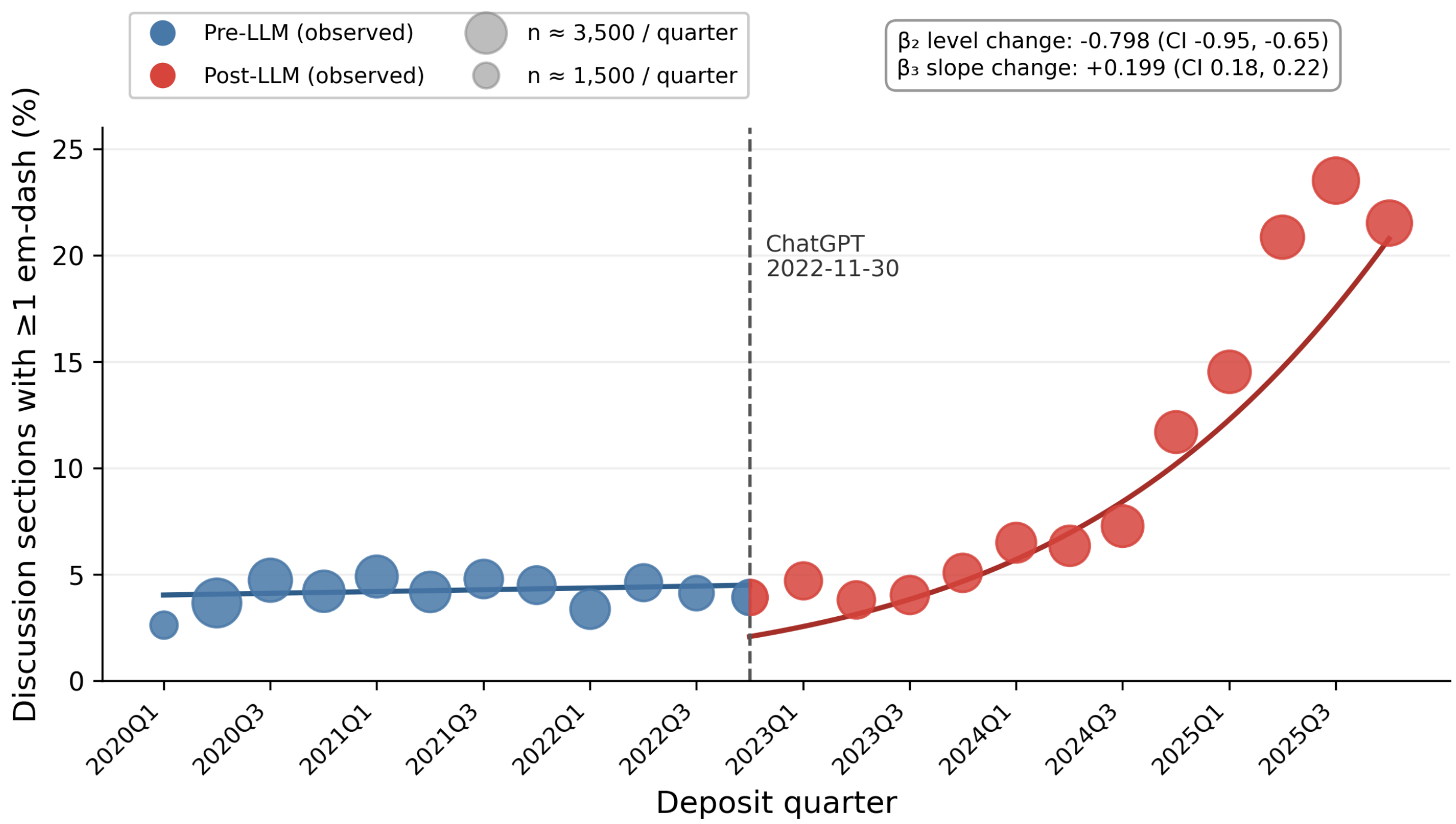

**Figure 3. Quarterly interrupted-time-series model, 24 quarters, breakpoint at the date of ChatGPT's release.** Points show the quarterly em-dash frequency, in blue before the LLM era and in red after it, with point size proportional to the number of preprints. The lines show the fitted segments before and after the breakpoint. The pre-breakpoint slope is essentially flat ($\beta_1$ = 0.010, $p$ = 0.27); the change of slope after the breakpoint is clearly positive ($\beta_3$ = +0.199, 95% CI 0.178–0.220), capturing the delayed acceleration rather than an immediate jump. The β coefficients are on the logit scale, not in percentage points. File: fig2_segmented_its_quarterly.pdf.

## Figure 4

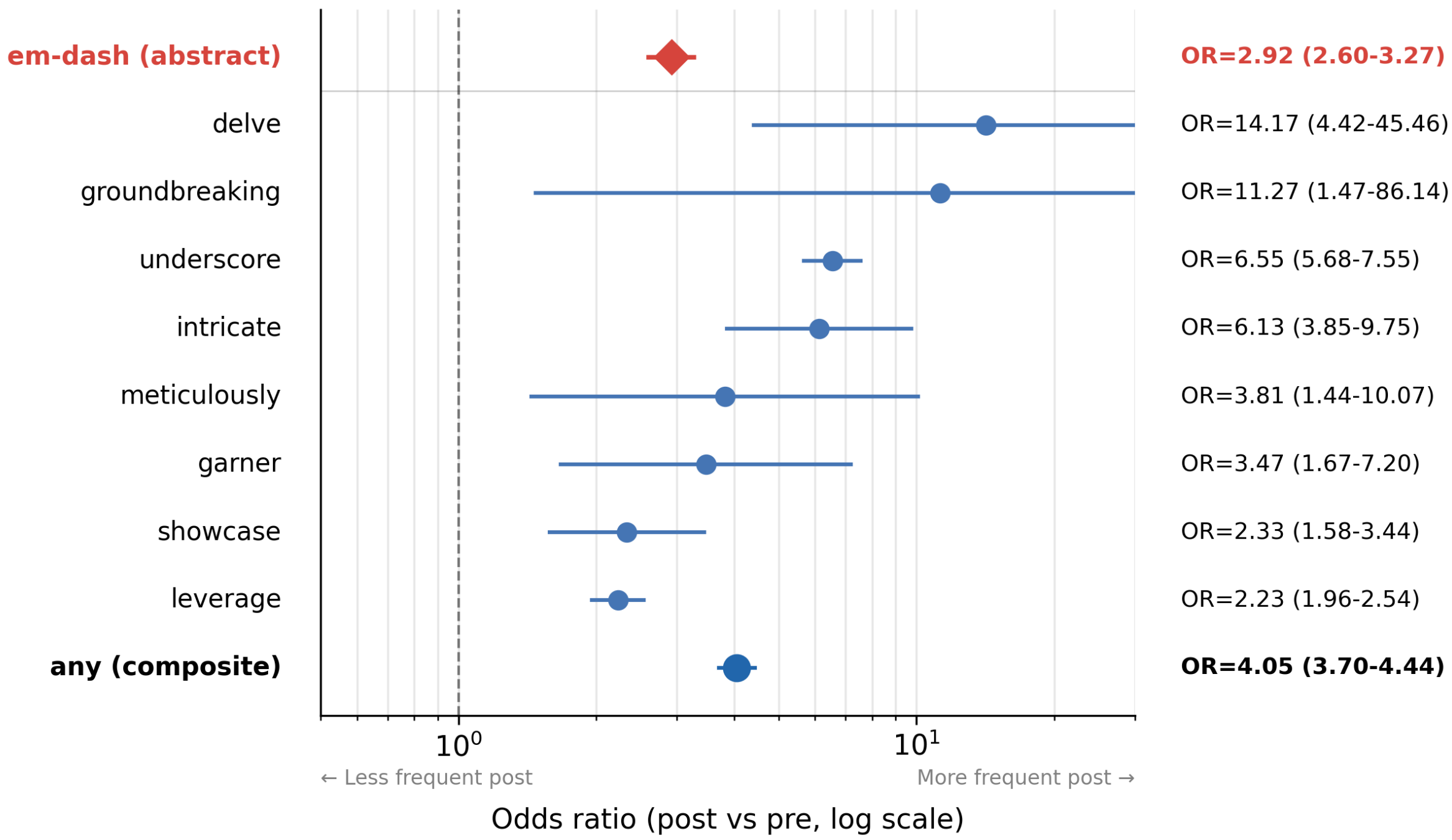

**Figure 4. Convergent lexical markers in abstracts, with the em-dash shown as a reference point: forest plot of odds ratios.** Per-word and composite odds ratios, post- versus pre-ChatGPT, for the eight most over-represented style words from Kobak et al. (2025), on a logarithmic scale. The composite odds ratio for any of the eight words is 4.05 (95% CI 3.70–4.44). The top row, marked with a red diamond, shows this study's own marker, em-dash presence, measured at the abstract level on the identical cohort for comparability, with an odds ratio of 2.92 (95% CI 2.60–3.27). The study's primary estimand is the Discussion-section em-dash (odds ratio 2.96), not this abstract-level reference value. This independent lexical signal moves in the same direction as the typographic marker. File: fig3_kobak_abstracts_forest.pdf.

**Figure 5**

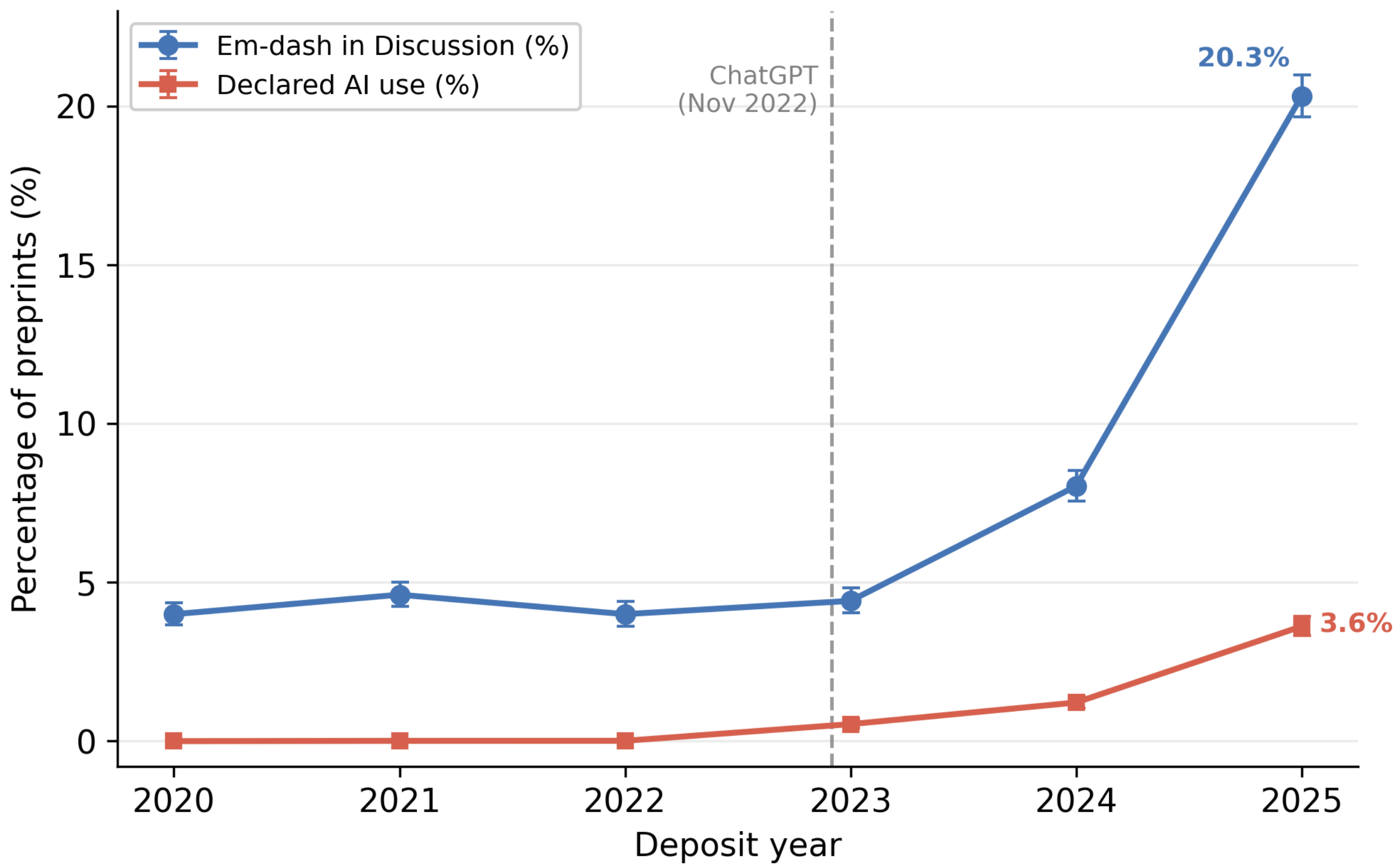

**Figure 5. Self-declared AI use and em-dash frequency rise in parallel after ChatGPT: exploratory analysis.** The percentage of first-version medRxiv preprints, by deposit year, whose Discussion contains at least one em-dash (blue circles), and the percentage carrying an explicit declaration of generative-AI use in preparing the manuscript (red squares), on a shared percentage axis, with 95% Wilson confidence intervals. The dashed line marks the public release of ChatGPT in November 2022. Both series are essentially flat before 2022 and rise in parallel thereafter: the em-dash to 20.3%, and declarations to 3.6% in 2025. These are two independent signals of the same change: one self-reported and one typographic. Exploratory, post-registration analysis; declarations were identified by an automated detector whose accuracy was assessed by an independent model acting as a judge (Claude Haiku, n = 2,162): precision 79.6%, recall 82.0%. File: fig4_declaration_vs_emdash.pdf.

---

# Declarations

**Pre-registration.** OSF doi:10.17605/OSF.IO/HFT8C (Stage 2 confirmatory; deposited before the confirmatory analysis was computed; under embargo until publication).

**Code and data.** Analysis code frozen at github.com/P-Czum/em-ergence-em-dash, commit 21b65fd. Processed per-paper measurements are archived on Zenodo (doi:10.5281/zenodo.20557419), released at publication. The medRxiv corpus is public external data (S3 Text-and-Data-Mining bucket) and is not redistributed. As a reproducibility check, all confirmatory, supporting, and sensitivity analyses were independently re-implemented from the pre-registered specification and reproduced the reported estimates.

**AI-assisted analysis disclosure.** AI tools (Claude, Anthropic) were used for code generation, statistical-analysis assistance, and editorial review. They did not formulate the research question, choose the corpus, or independently make scientific decisions. The author takes full responsibility for the content.

**Competing interests.** The author declares no competing financial interests. Non-financial interests: the author is the founder of the Polish Association for Artificial Intelligence in Medicine (inteligencja.org.pl) and advocates the use of artificial intelligence, including large language models, to improve human health, including through the support of science. The author is also developing a free, non-commercial LLM-based tool that summarises medical literature, with no plans for commercialisation; that work is the source of the author's familiarity with programmatic retrieval of biomedical articles via public APIs, which informed the technical approach taken here.

The tool itself was not used to obtain or process the data in this study, is not a subject of the study, and played no role in its design, analysis, or interpretation.

**Funding.** No external funding supported this work; all costs of the study were borne by the author.